\documentclass[sigconf]{acmart}

\AtBeginDocument{%
  }

\setcopyright{acmlicensed}
\copyrightyear{2026}
\acmYear{2026}
\acmDOI{XXXXXXX.XXXXXXX}

\acmConference[CAIS '26]{ACM Conference on AI and Agentic Systems}{May 26 -- 29, 2026}{San Jose, CA}

\usepackage{graphicx}
\usepackage{listings}
\usepackage{xcolor}
\usepackage{booktabs}

\lstdefinestyle{yaml}{
  basicstyle=\color{black}\scriptsize\ttfamily,
  rulecolor=\color{black},
  string=[s]{'}{'},
  stringstyle=\color{blue},
  comment=[l]{:},
  commentstyle=\color{black},
  morecomment=[l]{-},
  frame=single,
  xleftmargin=0.15cm,
  xrightmargin=0.15cm,
  breaklines=true
}

\begin{document}

\title{Token Management in Multi-Tenant AI Inference Platforms}
\author{William J. Cunningham}
\email{william.cunningham@datarobot.com}
\orcid{0000-0003-1161-5679}
\affiliation{%
  \institution{DataRobot}
  \city{Boston}
  \state{Massachusetts}
  \country{USA}
}

\begin{abstract}
Multi-tenant AI inference platforms must balance resource utilization against service-level guarantees under variable demand. Conventional approaches fail to achieve this balance: dedicated endpoints strand capacity on idle models, while rate limits ignore the heterogeneous cost of inference requests. We introduce \emph{token pools}, a control-plane abstraction that represents inference capacity as explicit entitlements expressed in inference-native units (token throughput, KV cache, concurrency). Unlike rate limits, which govern request admission without regard to execution cost, token pools authorize both admission and autoscaling from the same capacity model, ensuring consistency between what is promised and what is provisioned. The abstraction captures burst modes across multiple dimensions invisible to conventional throttling. Dynamic per-entitlement limits on each burst dimension enable fine-grained control over resource consumption while permitting work-conserving backfill by low-priority traffic. The design supports priority-aware allocation, service tiers with differentiated guarantees, and debt-based fairness mechanisms, all without modifying the underlying inference runtime or cluster scheduler. In experiments on a Kubernetes cluster with vLLM backends, token pools maintain a bounded P99 latency for guaranteed workloads during overload by selectively throttling spot traffic, while a baseline without admission control experiences unbounded latency degradation across all workloads. A second experiment demonstrates debt-based fair-share convergence among elastic workloads with heterogeneous SLO requirements during capacity scarcity.
\end{abstract}




\received{27 February 2026}

\maketitle

%==============================================================================
\section{Introduction}
%==============================================================================

Large language models have transformed enterprise computing, yet the infrastructure required to serve them efficiently remains immature. The autoregressive nature of transformer-based generation~\cite{vaswani2017attention} creates unique scheduling challenges: a single inference request may consume gigabytes of GPU memory for key-value (KV) cache, occupy decode slots for seconds during reasoning~\cite{openai2024o1,deepseek2025r1}, and exhibit order-of-magnitude variation in resource consumption depending on prompt length and output complexity~\cite{kwon2023vllm}. Conventional scheduling abstractions, designed for CPU-bound microservices or batch compute~\cite{verma2015borg,hindman2011mesos}, fail to capture these dynamics.

The dominant approaches to multi-tenant inference each sacrifice a critical property. \emph{Dedicated endpoints} provision GPU instances per model or tenant, achieving strong isolation but stranding capacity when models are idle; the long tail of infrequently-used models exacerbates this inefficiency. \emph{Rate limits} offer an alternative but govern request admission without regard to execution cost: commercial APIs enforce tokens-per-minute quotas~\cite{openai2024ratelimits} that treat all tokens equally, yet two requests may differ by an order of magnitude in GPU time and KV cache consumption depending on sequence length and model architecture~\cite{wilkins2024workload}. Neither approach supports work-conservation, so idle capacity cannot be borrowed by other tenants. Both also fail under the bursts that characterize inference traffic, including prompt-length bursts that exhaust KV cache, output-length bursts from reasoning models, and concurrent session bursts from parallel workloads.

Recent systems have addressed specific bottlenecks. Continuous batching~\cite{yu2022orca} and PagedAttention~\cite{kwon2023vllm} improve throughput by packing requests dynamically and managing KV cache as virtual memory. FlashAttention~\cite{dao2022flashattention} reduces memory bandwidth bottlenecks through IO-aware computation. Disaggregated architectures~\cite{zhong2024distserve,patel2024splitwise} separate prefill and decode phases onto specialized hardware, improving latency isolation. Mooncake~\cite{qin2024mooncake} elevates KV cache to a first-class schedulable resource, achieving substantial throughput gains by treating cache placement as an explicit optimization target. Speculative decoding~\cite{leviathan2023speculative} accelerates generation by predicting multiple tokens per forward pass. These advances improve execution efficiency but do not address \emph{multi-tenant capacity allocation}: who gets how much capacity, under what guarantees, and what happens when demand exceeds supply.

We propose \emph{token pools} as an inference-native resource abstraction for multi-tenant capacity management. A token pool exposes an autoscaling group of GPU workers in terms of three schedulable resources: token throughput (tokens/second), KV cache capacity (bytes), and request concurrency (active sequences). Tenants request \emph{entitlements} to portions of pool capacity; entitlements authorize both API admission and autoscaling decisions from a unified capacity model. Service classes (dedicated, guaranteed, elastic, spot, preemptible) determine protection ordering during contention, while a debt-based priority mechanism drives fair-share convergence over time.

The key insight is that admission control at the API boundary, rather than at the container or GPU level, enables fine-grained resource management without modifying inference runtimes. When a request arrives, the system evaluates whether the tenant's entitlement can fund the request under current conditions. Burst capacity is satisfied by reallocating unused tokens before triggering scaling; low-priority work is throttled to protect high-priority workloads. This approach closes the gap between admission decisions (which occur at the gateway) and execution costs (which are determined by the inference runtime).

Token pools operate as a control-plane layer above existing infrastructure. The underlying inference runtime (vLLM, TensorRT-LLM) and cluster scheduler remain unmodified; the contribution is not a new execution engine but a resource abstraction that mediates access to existing engines. This design principle enables adoption without replacing mature, well-understood backends.

Our contributions are: (1) formalization of token pools as inference-native schedulable resources, decomposing capacity into throughput, KV cache, and concurrency with a priority mechanism that combines service class, SLO urgency, burst history, and accumulated service debt; (2) a system architecture that repurposes the Kubernetes scheduler for token capacity admission via virtual nodes; and (3) experimental evidence that token pools maintain bounded latency for protected workloads during overload (sub-1.2s P99 vs 19+s baseline) while enabling fair-share convergence among elastic workloads with heterogeneous SLO requirements.

%==============================================================================
\section{Related Work}
%==============================================================================

\paragraph{Inference Serving Systems.}
Orca~\cite{yu2022orca} introduced iteration-level scheduling and continuous batching for transformer inference, achieving a 36$\times$ throughput improvement over static batching by scheduling at the granularity of individual iterations rather than entire requests. vLLM~\cite{kwon2023vllm} extended this with PagedAttention, managing KV cache as virtual memory to eliminate fragmentation and enable flexible sharing across requests. SGLang~\cite{zheng2024sglang} further optimizes for structured generation programs with RadixAttention for prefix sharing. Kthena~\cite{kthena2026} provides Kubernetes-native inference orchestration with KV cache-aware routing and topology-aware scheduling, but does not formalize capacity in inference-native units or provide debt-based compensatory allocation. Earlier systems like Clipper~\cite{crankshaw2017clipper} and INFaaS~\cite{romero2021infaas} provided model-agnostic serving abstractions, while Clockwork~\cite{gujarati2020clockwork} achieved predictable latency through centralized scheduling of GPU kernels. These systems are tenant-agnostic at the execution layer, delegating multi-tenant policy to external components. Token pools provide that policy layer, enabling differentiated capacity allocation based on tenant identity and service requirements.

\paragraph{Disaggregated Inference.}
DistServe~\cite{zhong2024distserve} and Splitwise~\cite{patel2024splitwise} separate prefill and decode phases onto different GPU pools, enabling independent scaling and reducing interference between compute-bound prefill and memory-bound decode. DistServe achieves up to 7.4$\times$ higher serving capacity by co-optimizing resource allocation for time-to-first-token (prefill) and time-per-output-token (decode) independently. Sarathi-Serve~\cite{agrawal2024sarathi} introduces chunked-prefills to balance the throughput-latency tradeoff. Mooncake~\cite{qin2024mooncake} takes disaggregation further, treating KV cache as a first-class resource with explicit placement and scheduling, achieving up to 5$\times$ capacity improvements for long-context workloads. NVIDIA Dynamo~\cite{dynamo2025} provides a production framework for disaggregated serving with latency-aware routing and KV cache management across multiple backends. llm-d~\cite{llmd2025} offers Kubernetes-native distributed inference with prefix-aware routing and variant autoscaling. These systems optimize execution efficiency but remain tenant-agnostic, delegating capacity allocation to external layers. Token pools complement disaggregated backends by adding service-class differentiation and fair-share mechanisms.

\paragraph{Quota and Entitlement Systems.}
Borg~\cite{verma2015borg} introduced quota as a long-term resource entitlement, with admission control rejecting jobs that exceed tenant quota at the requested priority level. Twine~\cite{tang2020twine} generalized this through dynamic entitlements that adjust machine allocations based on workload fluctuations, treating entitlements as ``pseudo-clusters'' with caps that can be reassigned in real time. Dominant Resource Fairness (DRF)~\cite{ghodsi2011drf} provides a theoretical foundation for fair allocation across multiple resource types, while Tetris~\cite{grandl2014tetris} extends this to bin-packing workloads. These systems allocate coarse-grained resources (machines, CPU cores) rather than inference-native units, and their fairness mechanisms operate at job granularity (minutes to hours) rather than request granularity (milliseconds). Token pools adapt entitlements to inference by expressing capacity in tokens/second and enforcing fairness per-request.

\paragraph{GPU Scheduling.}
GPU cluster schedulers have evolved from training-focused systems like Gandiva~\cite{xiao2018gandiva}, which time-slices jobs at iteration boundaries, to fairness-oriented schedulers like Tiresias~\cite{gu2019tiresias} and Themis~\cite{mahajan2020themis}. Fine-grained GPU sharing systems such as Salus~\cite{yu2020salus} enable fast job switching through memory sharing primitives, while Baymax~\cite{chen2016baymax} orchestrates latency-critical inference alongside best-effort background jobs. KAI-Scheduler~\cite{kaischeduler2025} extends Kubernetes with hierarchical GPU quotas, gang scheduling, and fair-share policies. These schedulers operate at pod granularity: workloads are containers requesting GPUs, and fairness governs pod placement. For inference workloads, this abstraction is limiting. A model server pod may be fully utilizing its GPU while starving high-priority requests because lower-priority requests arrived first. The core issue is that GPUs are bound to long-running model servers, and contention occurs at the request level rather than the pod level. Token pools relocate the control point from pod placement to request admission, enabling sub-second priority decisions and graceful degradation rather than disruptive preemption.

%==============================================================================
\section{Token Pool Formalism}
%==============================================================================

\subsection{Resource Model}

We consider a multi-tenant inference platform where tenants share a pool of GPU-backed compute resources. Each tenant holds one or more \emph{entitlements} to capacity in a token pool; entitlements are the unit of resource allocation and are indexed by $e$. The primary schedulable resources are:

\begin{itemize}
\item \textbf{Token throughput} $\lambda$ (tokens/second): the rate of token production, directly bounding service capacity and GPU time consumption. For a request with input length $n_{\text{in}}$ and output length $n_{\text{out}}$, the nominal token cost is $n_{\text{in}} + n_{\text{out}}$. Prefill and decode contribute differently to GPU utilization, but this approximation suffices for admission control.
\item \textbf{KV cache capacity} $\chi$ (bytes): the memory constraint for attention state. For a transformer with $L$ layers, $H_{kv}$ key/value heads, head dimension $d_h$, and element size $b$ bytes, each token requires $c = 2LH_{kv}d_hb$ bytes for keys and values (ranging from tens of KB for models with grouped-query attention to several MB for large dense models). A sequence with context length $S$ consumes $\chi = Sc$ bytes. For long-context models, KV cache is often the limiting factor.
\item \textbf{Concurrency} $r$ (active sequences): the number of simultaneously active inference sequences whose KV state is resident on decode GPUs. (We assume GPU-resident KV cache throughout; paged or offloaded cache is discussed in \S\ref{sec:discussion}.) This bounds batch size and decode slot competition. For a GPU with KV budget $\chi_{\text{gpu}}$, maximum concurrency is $r_{\max} = \lfloor \chi_{\text{gpu}} / (Sc) \rfloor$.
\end{itemize}

These three resources impose complementary constraints. Token throughput determines how fast work advances; KV cache determines how much state may exist simultaneously; concurrency determines how many sequences compete for decode slots. Conventional rate limits~\cite{raghavan2007cloud} govern only the first dimension, missing the latter two entirely. Models with extended context windows~\cite{brown2020gpt3,touvron2023llama} and mixture-of-experts architectures~\cite{jiang2024mixtral} amplify these distinctions: long contexts stress KV cache while sparse activation reduces compute per token.

Each entitlement $e$ specifies baseline allocations $(\lambda_e, \chi_e, r_e)$ representing the tenant's right to consume inference capacity. The aggregate entitled capacity of pool $p$ is $\sum_{e \in \mathcal{E}_p} \lambda_e$, where $\mathcal{E}_p$ denotes the set of entitlements bound to the pool. Let $\hat{\lambda}_e$ denote the actual allocation to entitlement $e$; feasibility requires $\sum_e \hat{\lambda}_e \leq \Lambda_p$, where $\Lambda_p$ is the pool's total throughput capacity derived from backend replica count and per-replica capacity.

\subsection{Service Classes}

Entitlements are assigned a service class $\kappa_e$ that governs baseline protection, burst rights, and eviction behavior. The class hierarchy defines a protection ordering: when reclaiming capacity, preemptible entitlements are evicted first, spot entitlements are throttled next, elastic entitlements are shrunk as needed, and dedicated/guaranteed entitlements are never touched.

\begin{table}[h]
\centering
\small
\begin{tabular}{@{}lcccc@{}}
\toprule
\textbf{Class} & \textbf{Baseline} & \textbf{Burst} & \textbf{Shrink} & \textbf{Weight} \\
\midrule
Dedicated & Reserved & Yes & Never & 1000 \\
Guaranteed & Reserved & No & Never & 1000 \\
Elastic & Time-Averaged & Yes & Yes & 100 \\
Spot & None & Yes & Yes & 1 \\
Preemptible & None & Yes & Evict & 0.1 \\
\bottomrule
\end{tabular}
\caption{Token pool service classes. Dedicated and guaranteed entitlements are never shrunk; elastic entitlements may be temporarily reduced below baseline; spot entitlements are throttled first; preemptible entitlements may be fully evicted.}
\label{tab:classes}
\end{table}

\textbf{Dedicated} entitlements receive reserved baseline allocation that is never reclaimed, even when idle, and may opportunistically consume idle capacity from other entitlements. \textbf{Guaranteed} entitlements also receive reserved baseline but cannot burst above it, providing predictable costs with traditional rate-limit semantics. \textbf{Elastic} entitlements receive baseline guarantees in aggregate over time via the debt mechanism: they may burst when capacity is available but may be shrunk below baseline when protected classes need capacity, accumulating debt that increases their priority for future allocation. \textbf{Spot} entitlements have no baseline guarantee, do not accumulate debt, and consume only surplus capacity; they are throttled first when utilization rises. The key distinction between Elastic and Spot is that Elastic workloads eventually receive compensatory allocation through the debt mechanism, while Spot workloads have no such guarantee. \textbf{Preemptible} entitlements may be fully evicted: active requests are terminated, KV cache is reclaimed, and the workload pod is killed.

\subsection{Priority and Debt Mechanism}

The priority weight $w_e$ combines class membership, SLO urgency, burst history, and service debt into a single scalar that drives all ordering decisions:
\begin{equation}
w_e = w_{\kappa_e} \cdot \left(1 + \alpha_{\text{slo}} \cdot \frac{\ell_e^*}{\bar{\ell}^*}\right)^{-1} \cdot \left(1 + \alpha_{\text{burst}} \cdot b_e\right)^{-1} \cdot \left(1 + \alpha_{\text{debt}} \cdot d_e\right)
\label{eq:priority}
\end{equation}
where $w_{\kappa_e}$ is the base weight for service class $\kappa_e$, $\ell_e^*$ is the SLO target (tighter targets yield higher priority), $\bar{\ell}^*$ is the pool-average SLO, $b_e$ is the burst intensity, and $d_e$ is the accumulated service debt. The coefficients $\alpha_{\text{slo}}$, $\alpha_{\text{burst}}$, and $\alpha_{\text{debt}}$ control sensitivity to each factor; typical values are $\alpha_{\text{slo}} = 2.0$, $\alpha_{\text{burst}} = 1.0$, and $\alpha_{\text{debt}} = 4.0$. The multi-order-of-magnitude gaps in class weights (1000 vs 100 vs 1 vs 0.1) ensure that class dominates other factors under normal conditions.

The \textbf{service debt} $d_e$ tracks whether an entitlement has been underserved over time, creating a feedback mechanism that drives allocation toward fair-share equilibrium. Define the instantaneous service gap as $g_e = (\lambda_e - \hat{\lambda}_e)/\lambda_e$, where positive $g_e$ indicates underservice (allocation below baseline) and negative $g_e$ indicates overservice (bursting above baseline). The debt accumulates via exponentially weighted moving average:
\begin{equation}
d_e(k) = \gamma_d \cdot d_e(k-1) + (1-\gamma_d) \cdot g_e(k)
\end{equation}
When $d_e > 0$, the multiplicative factor $(1 + \alpha_{\text{debt}} \cdot d_e)$ increases priority, making the system more likely to allocate resources to underserved entitlements. When $d_e < 0$ (accumulated credit from overservice), priority decreases, making the entitlement more likely to yield resources. This drives fair-share convergence: entitlements that have been temporarily shrunk accumulate debt that increases their priority until they receive compensatory allocation.

\begin{figure*}[!ht]
\centering
\includegraphics[width=\textwidth]{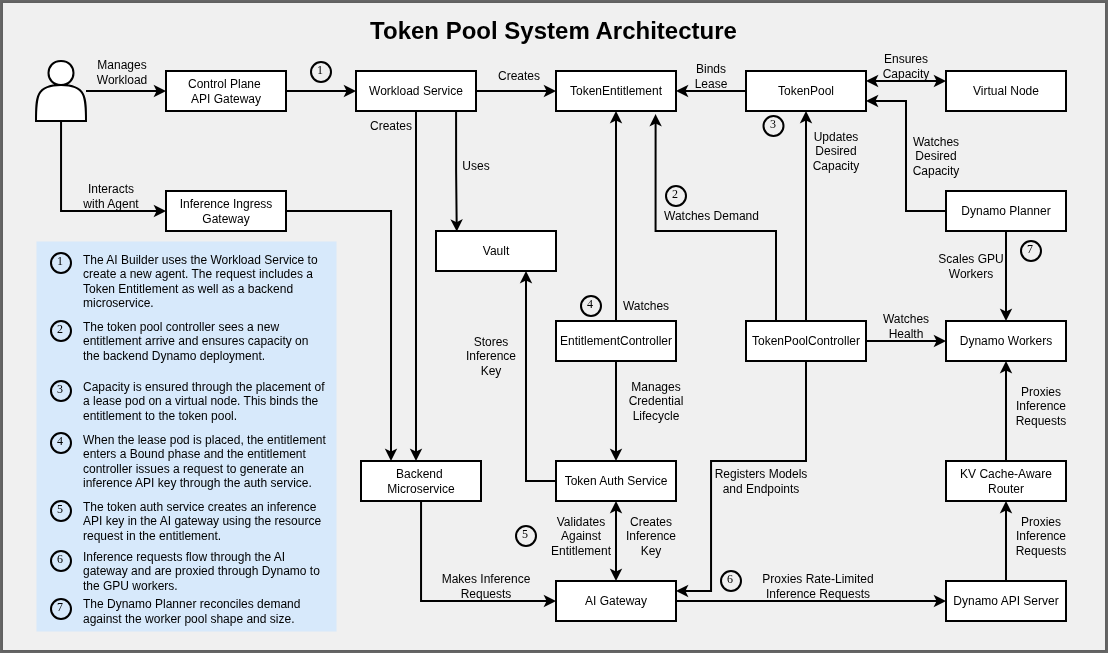}
\caption{System architecture. The TokenPool controller aggregates demand from entitlements and manages backend capacity while the Virtual Node Provider projects pool capacity into Kubernetes extended resources. AI Workloads access token-based compute resources through the AI Gateway using an inference key mapped to a particular entitlement. As demand fluctuates, the Dynamo planner reacts and scales GPU workers in order to meet service level objectives across tenants.}
\label{fig:arch}
\end{figure*}

The debt mechanism serves a role analogous to proportional-integral control: the service gap $g_e$ is the proportional term (instantaneous error), while the accumulated debt $d_e$ is the integral term (cumulative error). The EWMA decay $\gamma_d$ acts as an anti-windup mechanism, bounding the integral to prevent overcorrection after transient disturbances.

The \textbf{burst intensity} $b_e$ tracks overconsumption above baseline across all resource dimensions:
\begin{equation}
\delta_e = \max\left(0, \frac{\hat{\lambda}_e}{\lambda_e} - 1\right) + \max\left(0, \frac{\hat{\chi}_e}{\chi_e} - 1\right) + \max\left(0, \frac{\hat{r}_e}{r_e} - 1\right)
\end{equation}
capturing throughput bursts (request-rate and output-length), KV cache bursts (prompt-length and duration), and concurrency bursts (parallel sessions). The EWMA update $b_e(k) = \gamma_b \cdot b_e(k-1) + (1-\gamma_b) \cdot \delta_e(k)$ ensures that brief bursts incur minimal penalty while sustained bursting progressively reduces priority, preventing any entitlement from monopolizing shared capacity indefinitely.

%==============================================================================
\section{System Architecture}
%==============================================================================

Having defined the resource model and priority mechanism, we now describe how these abstractions are realized in a Kubernetes-native architecture. Figure~\ref{fig:arch} illustrates the system components and their interactions.

The central architectural challenge is bridging two resource models: token pools allocate capacity in inference-native units, while Kubernetes schedules containers to nodes based on CPU, memory, and extended resources. We resolve this through \emph{virtual nodes}: synthetic node objects that represent token pool capacity rather than physical machines.

\subsection{Virtual Node Abstraction}

For each TokenPool, the Virtual Node Provider creates a corresponding virtual node that advertises extended resources mirroring the pool's capacity: token throughput and KV cache in GiB. When the TokenEntitlement controller creates a \emph{virtual lease pod} requesting specific token resources, the Kubernetes scheduler evaluates whether the virtual node has sufficient allocatable capacity. If capacity is insufficient, the pod remains pending and the entitlement enters the Degraded state. If capacity is available, the scheduler binds the pod to the virtual node.

The lease pod consumes no actual compute; it exists solely to occupy capacity on the virtual node, preventing other entitlements from claiming the same resources. This repurposes the Kubernetes scheduler as the admission control mechanism for token capacity, inheriting its consistency guarantees and race-condition handling.

\subsection{Custom Resources}

Token pools and entitlements are represented as Kubernetes custom resources, enabling declarative configuration and integration with existing GitOps workflows. A \textbf{TokenPool} resource defines a logical capacity pool bound to a model backend with autoscaling bounds:

\begin{lstlisting}[style=yaml]
apiVersion: tokens.datarobot.com/v1alpha1
kind: TokenPool
metadata:
  name: qwen3-8b
spec:
  model: Qwen/Qwen3-8B
  backendRef:
    kind: DynamoGraphDeployment
    name: qwen3-8b-backend
  scaling:
    minReplicas: 1
    maxReplicas: 10
\end{lstlisting}

A \textbf{TokenEntitlement} resource grants a tenant access to pool capacity with specified resource limits and QoS parameters:

\begin{lstlisting}[style=yaml]
apiVersion: tokens.datarobot.com/v1alpha1
kind: TokenEntitlement
metadata:
  name: sample-entitlement
spec:
  tenantId: "3ed0feec"
  poolRef:
    name: qwen3-8b
  qos:
    serviceClass: guaranteed
    sloTargetMs: 200
  resources:
    tokensPerSecond: 100
    kvCacheGiB: 2
    concurrency: 4
\end{lstlisting}

\subsection{Admission Control}

The auth service intercepts every request before it reaches the backend. On each request, it validates the API key against an entitlement, retrieves the entitlement's current state from Redis, and makes an admission decision based on multiple factors:

\begin{enumerate}
\item \textbf{Entitlement state}: The entitlement must be in Bound state (not Pending, Degraded, or Expired). Checks are evaluated in order; a failing check short-circuits evaluation.
\item \textbf{Output length bound}: If the request omits \texttt{max\_tokens}, a configurable default is applied for capacity planning purposes.
\item \textbf{Concurrency limit}: The entitlement's in-flight request count must be below its concurrency limit $r_e$.
\item \textbf{Token budget}: The request's token budget (input tokens plus \texttt{max\_tokens}) must fit within the entitlement's remaining throughput allocation.
\item \textbf{Pool contention}: If the pool is contended, the request's priority $w_e$ must exceed the pool's admission threshold.
\end{enumerate}

Under contention, the admission threshold is set to the minimum priority among currently-admitted requests. Requests with priority below this threshold are rejected with HTTP 429 and a \texttt{Retry-After} header, allowing clients to back off gracefully. Requests with sufficient priority are admitted. This selective throttling directs capacity scarcity toward lower-priority workloads while protecting higher-priority entitlements.

The auth service maintains per-entitlement state in Redis for low-latency access: current in-flight count, burst intensity $b_e$, accumulated debt $d_e$, and effective allocation. These values are updated on every request completion via the gateway's callback mechanism: when a request finishes, the gateway posts actual token consumption and latency to the auth service, which updates the burst and debt terms based on observed resource usage. This closes the loop between admission (which occurs before execution) and cost accounting (which occurs after execution). The gateway integration for testing uses LiteLLM~\cite{litellm2024}, an open-source proxy that provides OpenAI-compatible endpoints with extensible authentication hooks.

%==============================================================================
\section{Evaluation}
%==============================================================================

We evaluate token pools through two experiments on a Kubernetes cluster with vLLM backends. The first validates cross-class protection: guaranteed workloads are insulated from spot traffic during overload. The second validates the debt-based priority mechanism: elastic workloads with different SLO requirements converge toward fair-share allocation during capacity scarcity.

\subsection{Experimental Setup}

Experiments run on a single-node cluster with one vLLM~\cite{kwon2023vllm} replica serving \texttt{nvidia/Qwen3-8B-NVFP4}, providing capacity for 16 concurrent sequences (hereafter ``slots'') at a total throughput of roughly 240 tokens/sec. We use a constrained setup on a DGX Spark for several reasons. First, token pools operate at the admission control layer, where behavior depends on the ratio of demand to capacity rather than absolute scale. The logic is scale-invariant by construction: it operates on ratios and priorities, not absolute counts. Production validation at larger scale remains future work. Second, a small, reproducible environment enables precise control over contention timing and eliminates confounding variables from distributed coordination, network partitions, or heterogeneous hardware. Third, the modest hardware requirements (a single GPU node) allow other researchers to reproduce these experiments without access to large-scale infrastructure. Production deployments would span multiple replicas and GPU types, but the admission control logic remains unchanged.

The token pool system comprises the TokenPool and TokenEntitlement controllers, a Redis-backed auth service for admission control, and LiteLLM as the API gateway. The baseline represents the absence of admission control: all requests are admitted regardless of capacity, and latency degrades for all workloads equally under overload. This baseline isolates the effect of admission control; comparison against token-based rate limits is discussed in Section~\ref{sec:discussion}. Traffic is generated using configurable load patterns; Experiment 1 uses 64-token input and output sequences, while Experiment 2 uses variable lengths (32--176 tokens) to differentiate workload characteristics.

\subsection{Experiment 1: Cross-Class Protection}

\emph{Scenario: ``Someone's batch job flooded the inference endpoint and our production latency spiked.''}

Three entitlements share a pool with 16 concurrent slots: \textsf{guaranteed-a} (6 slots), \textsf{spot-b} (10 slots), and \textsf{guaranteed-c} (6 slots, joining at $t$=30s and departing at $t$=60s). During Phase 2 (30--60s), total demand is 22 slots against 16 available, creating 38\% overload.

\begin{figure}[t]
\centering
\includegraphics[width=\columnwidth]{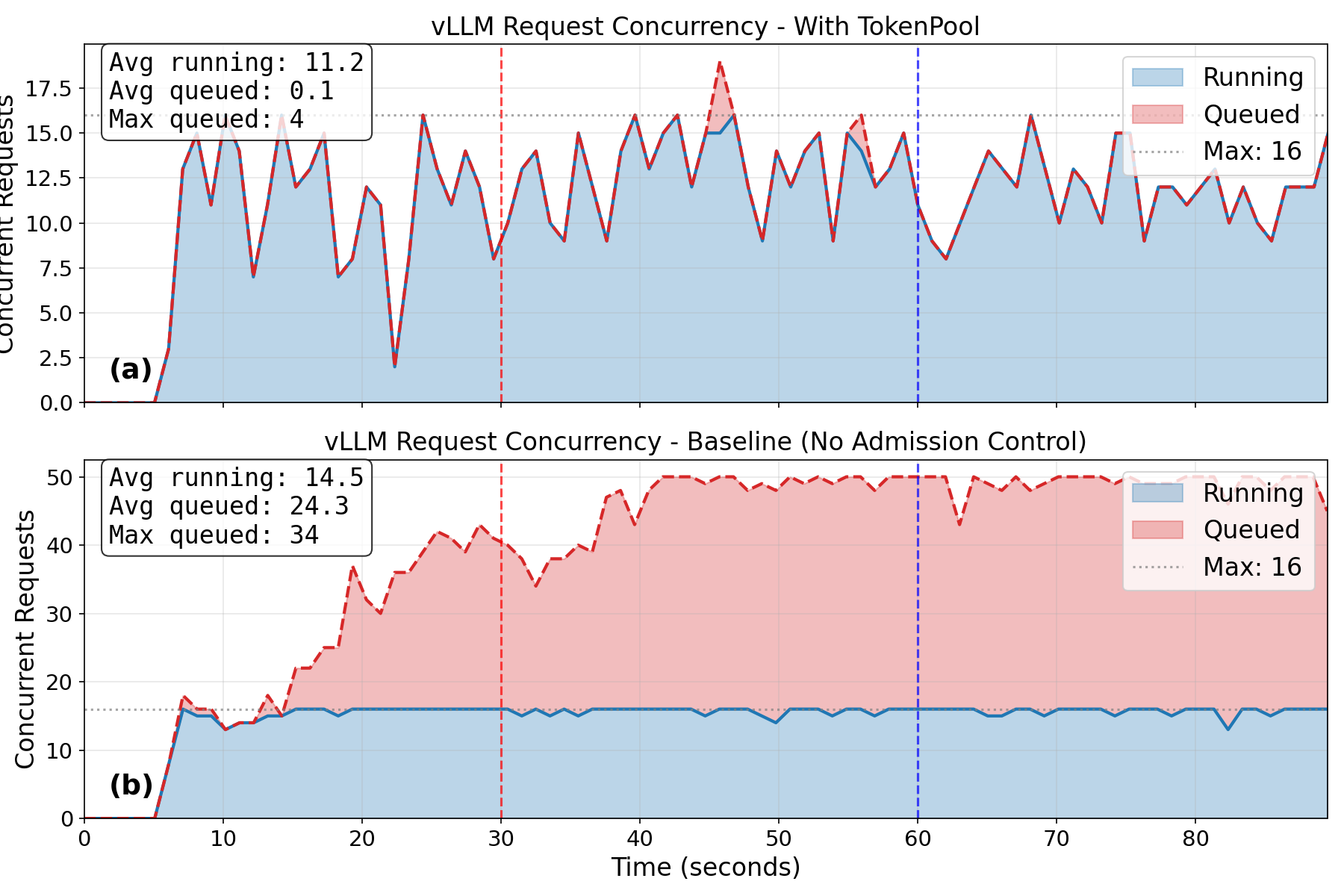}
\caption{Request queue depth during overload. (a) With token pools, running requests remain at capacity while the waiting queue stays empty; excess spot requests receive HTTP 429 responses. (b) Without admission control, the request queue grows to 34 requests, leading to sustained latency degradation.}
\label{fig:noisy-neighbor}
\end{figure}

\begin{figure}[t]
\centering
\includegraphics[width=\columnwidth]{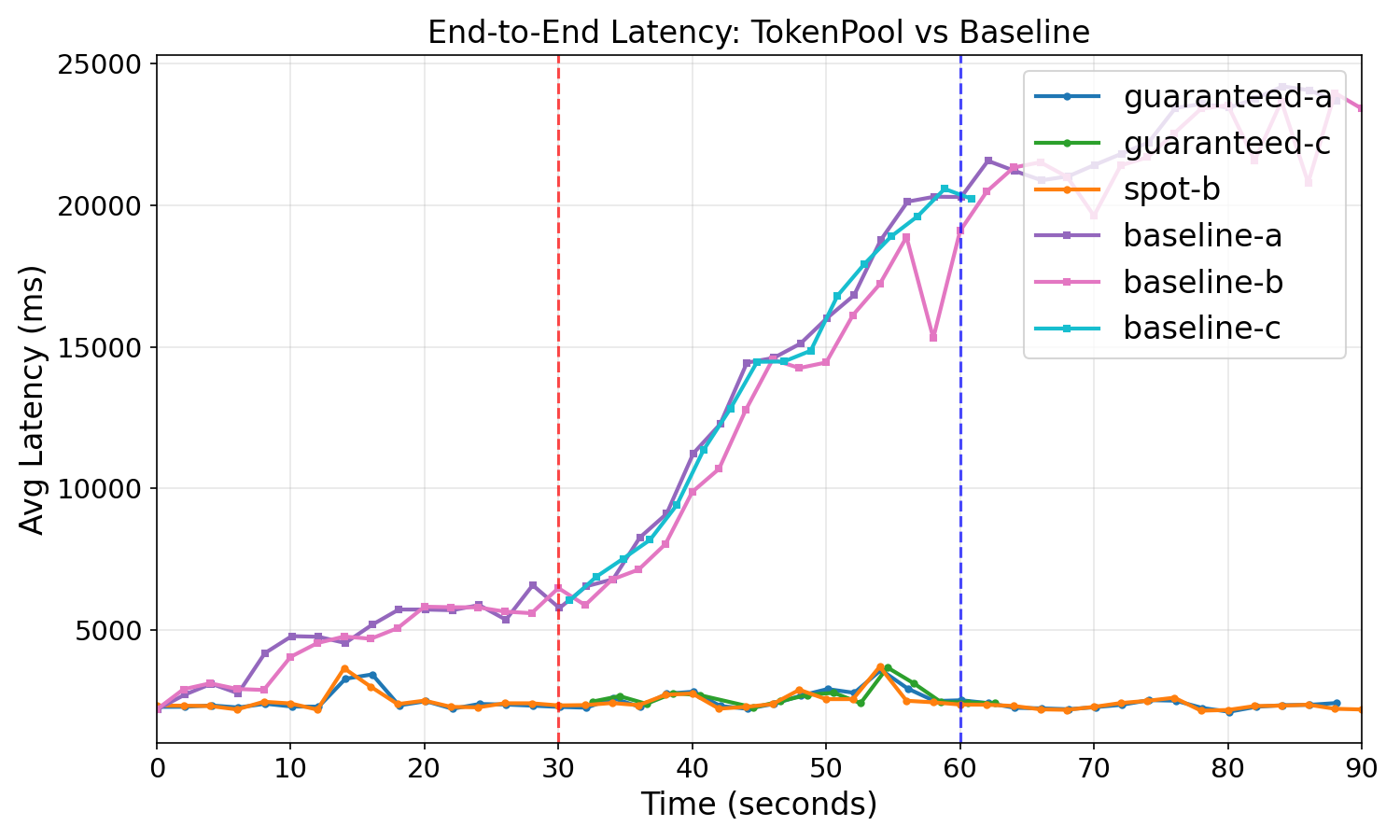}
\caption{End-to-end request latency. Token pools maintain bounded latency by rejecting excess spot requests; the baseline experiences unbounded latency growth as the queue deepens.}
\label{fig:latency-comparison}
\end{figure}

\begin{figure}[t]
\centering
\includegraphics[width=\columnwidth]{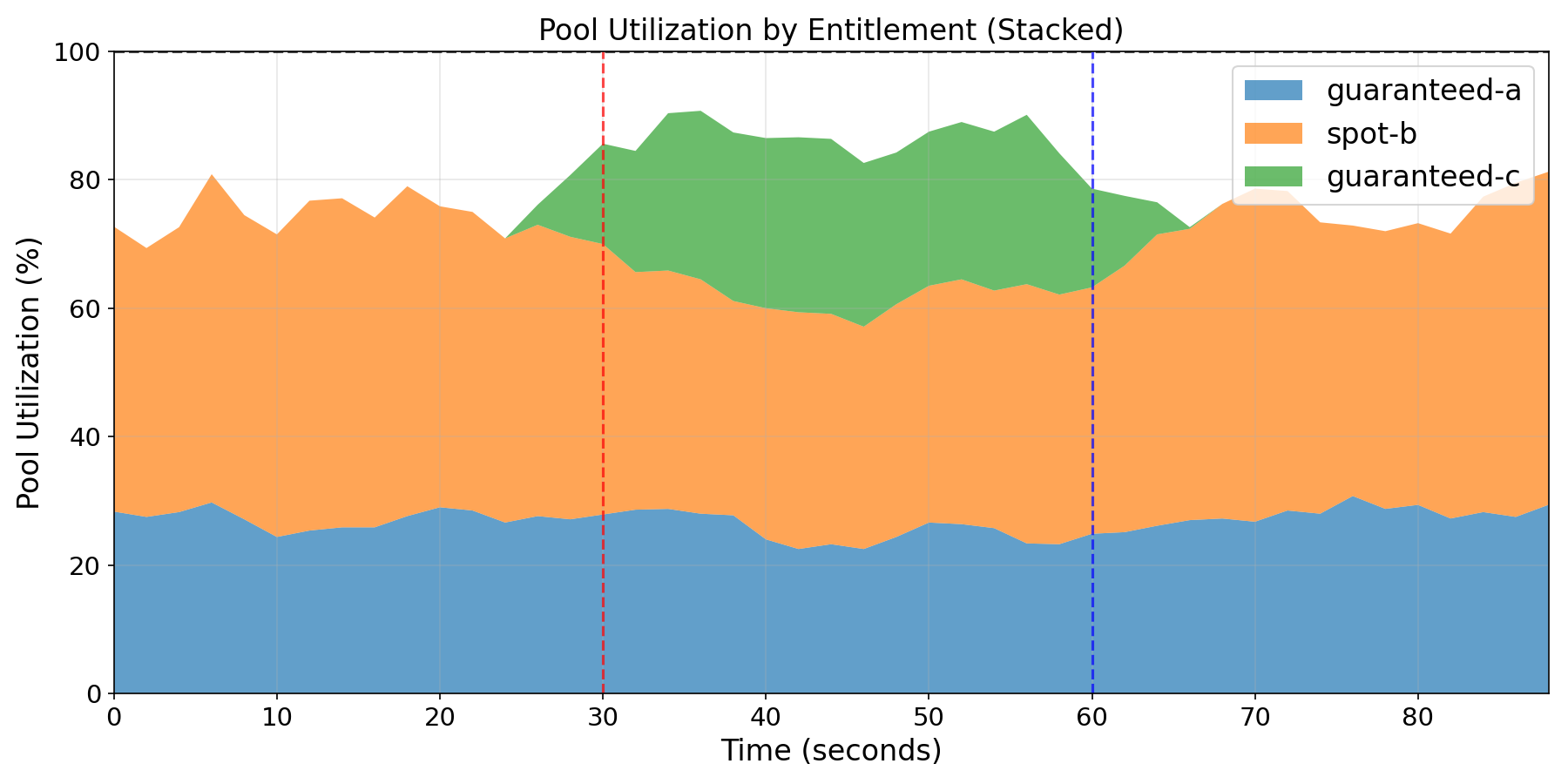}
\caption{Pool slot utilization by entitlement. Guaranteed workloads maintain their allocations while spot is squeezed during overload. Spot recovers immediately when guaranteed-c departs.}
\label{fig:pool-utilization}
\end{figure}

Figure~\ref{fig:noisy-neighbor} shows that with token pools, running requests remain at capacity while the waiting queue stays empty: excess spot requests receive HTTP 429 responses with \texttt{Retry-After} headers. Without admission control, the waiting queue grows to 34 requests, and latency grows continuously as the queue deepens. Figure~\ref{fig:latency-comparison} shows the latency impact directly: token pools maintain sub-1.2 second P99 TTFT for guaranteed workloads throughout all phases, while the baseline degrades to 19+ seconds by the end of Phase 2.

Figure~\ref{fig:pool-utilization} shows how pool capacity is allocated among entitlements. In Phase 1, guaranteed-a and spot-b share the 16-slot pool. When guaranteed-c joins at $t$=30s, the pool becomes fully subscribed with guaranteed workloads, and spot-b's share is squeezed to near zero. When guaranteed-c departs at $t$=60s, spot-b immediately recovers its share. The stacked areas sum to approximately 100\% throughout, confirming that the pool remains fully utilized while admission control redistributes capacity based on service class priority.

The key finding is that admission control at the API boundary prevents queue buildup that would otherwise degrade all workloads. Spot requests that cannot be served immediately are rejected with retry hints, allowing clients to back off gracefully. Guaranteed requests are never rejected (within their concurrency limits), and their latency remains bounded regardless of spot traffic patterns.

\subsection{Experiment 2: SLO-Aware Fair Share}

\emph{Scenario: ``A GPU node fails during peak hours. Two production services share the surviving capacity: a latency-critical coding assistant and a batch synthetic data pipeline. After recovery, an analytics report generator joins to diagnose what occurred.''}

Three elastic entitlements participate: \textsf{elastic-copilot} (5 slots, 500ms SLO) is a developer coding assistant requiring fast responses; \textsf{elastic-synth} (5 slots, 30s SLO) is a batch synthetic data pipeline tolerant of delay; and \textsf{elastic-reports} (5 slots, 5s SLO) is an analytics dashboard that joins at $t$=210s after recovery to generate incident reports. Phase 2 (30--120s) simulates an outage that halves capacity to 8 slots.

\begin{figure*}[pt]
\centering
\includegraphics[width=0.9\textwidth]{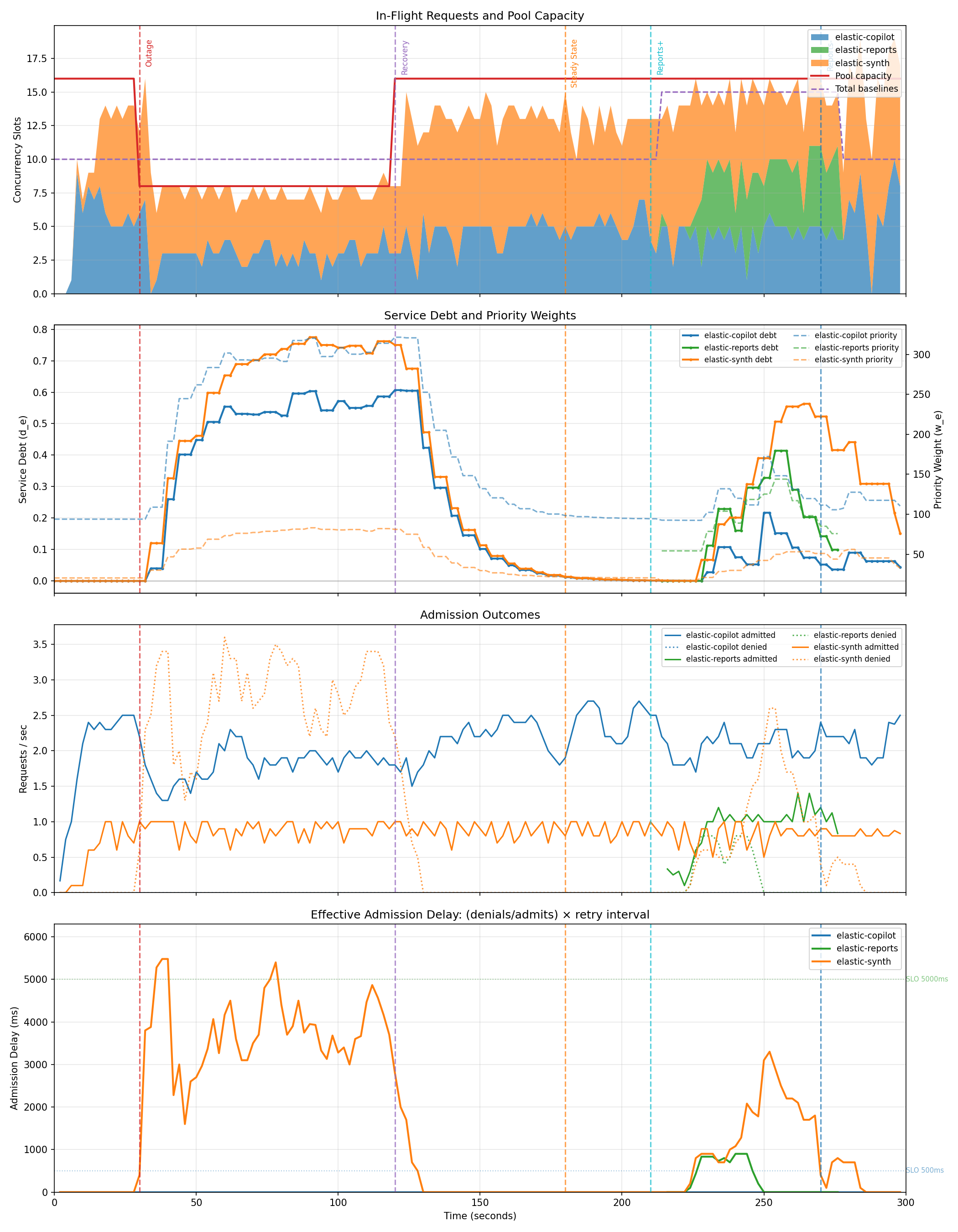}
\caption{Priority dynamics during capacity scarcity and recovery. Four panels: (1) in-flight requests with pool capacity; (2) service debt and priority weight; (3) admission and denial rates; (4) effective admission delay. Copilot (tight SLO) receives preferential admission while synth (loose SLO) absorbs throttling. Debt accumulates during underservice and decays during recovery.}
\label{fig:fair-share-dynamics}
\end{figure*}

With $\alpha_{\text{slo}} = 2.0$ and pool-average SLO $\bar{\ell}^* = 15{,}250$\,ms, the priority formula yields $w_{\text{copilot}} \approx 93.8$ and $w_{\text{synth}} \approx 20.3$. This 4.6$\times$ priority difference reflects copilot's tighter latency requirement.

Figure~\ref{fig:fair-share-dynamics} shows the priority dynamics over the 300-second experiment. Panel 1 shows total in-flight requests dropping to match reduced capacity during the outage, with copilot maintaining 5--7 slots while synth is compressed to 2--3. Panel 3 shows the mechanism: copilot receives zero low-priority denials throughout the outage, while synth accumulates 317. When the pool is full, the auth service compares each arriving request's priority against the admission threshold; copilot's 4.6$\times$ higher priority consistently exceeds this threshold while synth's does not.

\begin{table}[h]
\centering
\small
\begin{tabular}{@{}lccc@{}}
\toprule
\textbf{Metric} & \textbf{Copilot} & \textbf{Synth} & \textbf{Reports} \\
\midrule
Successful requests & 700 & 381 & 60 \\
Low-priority denials & 0 & 317 & 22 \\
Peak debt $d_e$ & 0.607 & 0.775 & 0.414 \\
\bottomrule
\end{tabular}
\caption{Admission outcomes for fair share experiment. Copilot receives zero low-priority denials while synth absorbs 317 during contention, demonstrating SLO-aware throttling.}
\label{tab:fair-share-results}
\end{table}

Panel 2 reveals the debt mechanism operating as designed. Both entitlements accumulate positive debt during the outage (both are underserved relative to their 5-slot baselines), but synth accumulates faster because it receives a smaller share of available capacity. Synth's debt peaks at $d_e \approx 0.775$ while copilot's peaks at $d_e \approx 0.607$. With $\alpha_{\text{debt}} = 4.0$, the debt term amplifies these into priority adjustments: at peak debt, synth's priority rises from 20.3 to approximately $20.3 \times (1 + 4.0 \times 0.775) = 83.2$, narrowing the priority gap from 4.6$\times$ to 3.9$\times$. While copilot retains higher priority throughout (as its SLO demands), synth's rising debt ensures it receives an increasing share of capacity as the outage persists, preventing starvation.

\begin{figure}[t]
\centering
\includegraphics[width=\columnwidth]{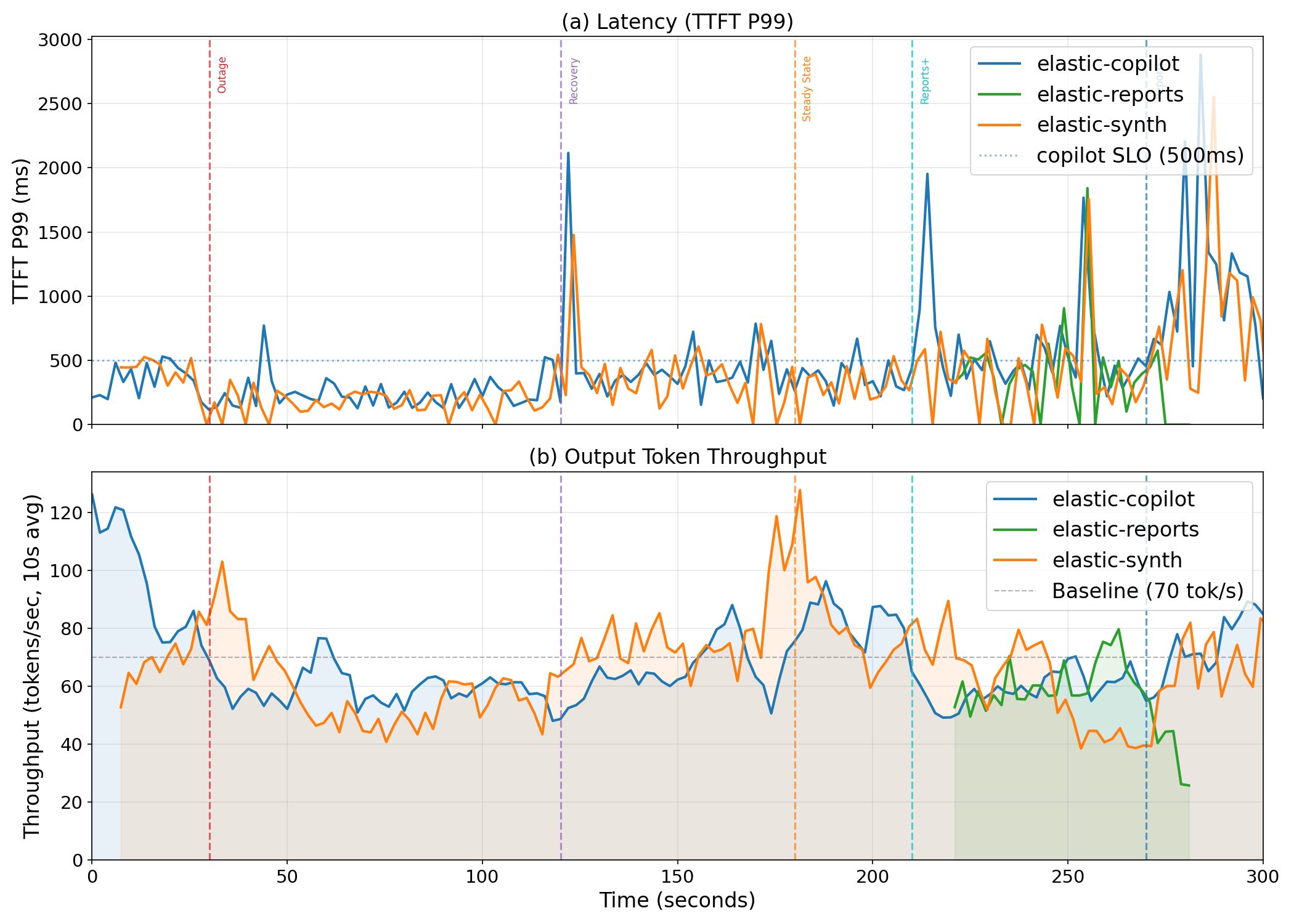}
\caption{Latency and throughput under dynamic priority allocation. (a) P99 TTFT remains largely within SLO for all entitlements despite contention, enabled by the short sequence lengths in this experiment. Exceptions reflect sudden capacity changes. (b) Output token throughput remains largely at the target value despite capacity and demand variations.}
\label{fig:fair-share-overview}
\end{figure}

When capacity is restored at $t$=120s, debt decays exponentially via the EWMA update with $\gamma_d = 0.7$, returning to near-zero within approximately 50 seconds. Priority weights return to their SLO-determined baselines. The transient TTFT spike at recovery (Figure~\ref{fig:fair-share-overview}) reflects a thundering-herd effect as retrying clients arrive simultaneously; in production, capacity restoration would be gradual.

When elastic-reports joins at $t$=210s with 5 slots of additional baseline demand, total baselines rise to 15 of 16 available slots. All three entitlements experience elevated denial rates during three-way competition. Reports enters with zero debt, competing based solely on its SLO term ($w_{\text{reports}} \approx 60$ for its 5s target). The debt mechanism ensures that newcomers are not privileged; they compete on equal footing with incumbents.

%==============================================================================
\section{Discussion}
\label{sec:discussion}
%==============================================================================

\paragraph{Summary of Findings.}
These experiments validate two core mechanisms at different levels of the service class hierarchy. Experiment 1 demonstrates cross-class protection: guaranteed workloads are insulated from spot traffic via selective throttling at the admission boundary, maintaining sub-1.2 second P99 TTFT while the baseline without admission control degrades to 19+ seconds. The 47\% spot throttle rate during overload represents the system correctly identifying which traffic to reject: spot requests are expendable while guaranteed requests are protected.

Experiment 2 demonstrates intra-class differentiation: among elastic workloads sharing the same service class, the SLO term in the priority formula directs throttling toward latency-tolerant workloads (synth, with 30s SLO), while the debt term accumulates compensatory priority for underserved entitlements. The priority gap narrowing from 4.6$\times$ to 3.9$\times$ during the outage shows debt working as intended: synth receives steadily increasing priority as its underservice accumulates, preventing starvation while still respecting the SLO-based ordering.

Together, the two experiments cover the primary allocation scenarios: protecting critical traffic from noisy neighbors (Experiment 1) and distributing scarce capacity fairly among peers with heterogeneous SLO requirements (Experiment 2).

\paragraph{Comparison to Static Rate Limits.}
Static per-entitlement rate limits could produce superficially similar results during steady state, but fail under dynamic conditions. Consider Experiment 2: with two entitlements configured for 5 slots each, rate limits would permit 10 concurrent requests. During the outage, only 8 slots are available, so the system must decide which entitlement to throttle and by how much. Static limits provide no basis for this decision; an operator would need to manually reconfigure limits during the outage and again after recovery. Token pools automate this by computing priority from SLO targets and accumulated debt, adapting to capacity changes without reconfiguration. The debt mechanism provides memory that static limits cannot express: an entitlement that was underserved during the outage receives compensatory priority during recovery.

\paragraph{Limitations.}
Experiments use a single vLLM replica; production deployments would involve multiple replicas and cross-replica coordination for pool-level state. The three-resource model assumes colocated prefill and decode; extending to disaggregated architectures would require phase-specific capacity terms and accounting for KV cache transfer bandwidth. Formal convergence analysis of the debt mechanism under arbitrary demand patterns is future work. The debt decay rate $\gamma_d$ was tuned manually; adaptive tuning based on workload characteristics could improve convergence. KV cache accounting assumes GPU-resident state; extending to paged or offloaded KV cache requires additional modeling.

Multi-model pools with sleep/wake transitions could reduce costs for long-tail models but remain unvalidated. Dedicated and preemptible service classes are defined but not exercised in these experiments.

%==============================================================================
\section{Conclusion}
%==============================================================================

Multi-tenant inference platforms face a fundamental tension: high utilization requires sharing capacity across tenants, but sharing without differentiation means that all tenants suffer equally when demand exceeds supply. Token pools resolve this tension by expressing capacity in inference-native units and enforcing priority-aware admission at the API boundary.

The key insight is that admission control belongs at the gateway, not the GPU scheduler. By the time a request reaches the inference runtime, the system has already committed resources; graceful degradation requires decisions before execution begins. Token pools provide the abstraction for these decisions: entitlements define what each tenant may consume, service classes determine protection ordering, and the debt mechanism drives fair-share convergence over time.

Experimental results validate both cross-class protection (guaranteed workloads maintain sub-1.2s P99 latency while spot traffic is throttled) and intra-class fairness (SLO-aware priority directs throttling toward latency-tolerant workloads while debt prevents starvation). The control-plane design operates above existing inference runtimes, enabling adoption without modifying production backends. As inference workloads grow in scale and heterogeneity, token pools offer a principled foundation for capacity management that respects both utilization goals and service-level commitments.

\begin{acks}
The author thanks R. Venkateswar, N. Purusothaman, D. Kothottil, and V. Bala for useful discussions, suggestions, and comments throughout the research and implementation of this work. This work was produced in part using AI models including Claude Opus 4.5 and GPT 5.2. All original ideas, prompts, and test cases, as well as the original draft manuscript, were produced by the author. Development was performed in part on a DGX Spark provided to DataRobot by NVIDIA.
\end{acks}

\bibliographystyle{unsrt}
\bibliography{paper}

\end{document}